\documentclass[a4paper]{article}

\pdfoutput=1

\usepackage{amssymb}
\usepackage{graphicx}
\usepackage[T1]{fontenc}
\usepackage{url}
\usepackage[american]{babel}
\usepackage{natbib}
\usepackage{hyperref}
\let\cite\citet

\RequirePackage{amsmath,amssymb,ifthen,url,graphicx,color,array,theorem}

\begin{document}

\title{Inhomogeneous CTMC Model of a Call Center with Balking and Abandonment}

\author{Maciej Rafal Burak\\
 \multicolumn{1}{p{.7\textwidth}}{\centering\emph{Applied Informatics,\\
West Pomeranian University of Technology\\
ul. Sikorskiego 37, Szczecin, Poland\\
\url{http://www.we.zut.edu.pl}}}}

\maketitle

\begin{abstract}
This paper considers a nonstationary multiserver queuing model with abandonment and balking for inbound call centers. We present a continuous time Markov chain (CTMC) model which captures the important characteristics of an inbound call center and obtain a numerical solution for its transient state probabilities using uniformization method with steady-state detection.

\end{abstract}

\section{Introduction}

The problem of managing operations of a telephone call center in an efficient way has a long history in the area of operational research and is a topic of current research in various disciplines (see e.g. \cite{Aksin_2007} or \cite{Gans_2003} for extensive overviews). From the modeling point of view they can be viewed as queuing systems. 

Such a queuing model can be described by a corresponding continuous time Markov chain (CTMC) whose steady-state distribution can be easily determined, either analytically - with the Erlang-C  formula for the simplest M/M/n model or with the Erlang-A formula for its version augmented with exponential patience time as proposed in \cite{Brown_2005}; or numerically for more complicated models  (as in \cite{Deslauriers_2007} or recently \cite{Phung_Duc_2014}). However, as real call centers are time inhomogenous, with varying arrival rates and changing number of servers - scheduled to meet the forecasted demand and in order to provide break time, stationary models cannot be applied directly. It is, therefore, common to use approximations, assuming the system being pointwise stationary. Examples of  such well established methods can be found e.g. in \cite{Green_2007coping}, \cite{Aksin_2007} or in \cite{Brown_2005}.
Unfortunately, stationary approximations are in many cases not adequate. For example, \cite{Deslauriers_2007} compared them with simulations based on real inbound call center data, with the conclusion that due to the nonstationarity only some of the performance measures can be estimated with satisfactory accuracy. Ingolfsson in \cite{Ingolfsson_2010} compared them with an inherently transient model and found their results significantly inaccurate or even entirely unreliable.
Despite this, their widespread use is commonly justified by simple implementation and low computational costs.

Many authors proposed to use simulation, which can achieve any desired accuracy. However, in order to achieve acceptable precision, very long computational times are needed, which makes it often impracticable for common applications like schedule planning.

An alternative approach, which is very effective in terms of the accuracy of the model, is to analyze transient CTMC using numerical methods, solving effectively their corresponding system of \emph{ordinary differential equations} (ODE's) as proposed in \cite{Ingolfsson_2007}, \cite{Bylina_2009} or by the author in \cite{burak2014multi}.

Other, less computationally intensive, analytical methods that can approximate such nonstationary systems more accurately than stationary models are \emph{closure approximations} and \emph{fluid and diffusion approximations},  discussed e.g. in \cite{Brown_2005}, \cite{Green_2007coping}, \cite{Czach_rski_2009} and   \cite{Czach_rski_2014} or, for the direct comparison of some examples of such methods with the numerical methods and stationary approximations, in \cite{Ingolfsson_2007}.

Although there is a number of papers dealing with the phenomena of customer balking and abandonment in multiserver queues (e.g. \cite{Brown_2005},\cite{Mandelbaum_2009},\cite{Whitt_2006},\cite{Artalejo_2009} or recently \cite{Phung_Duc_2014}), they concentrate on stationary models or approximations. To the best of our knowledge, an inherently transient CTMC model dealing with both balking and abandonment of a call center, has never been investigated.

The main objective of this work is to model such non-stationary systems, using transient analysis of corresponding CTMC, in a reliable and precise way, with computational efficiency enabling its use for practical applications -- in particular, as a much more accurate replacement to the Erlang-C and Erlang-A formulas, used by practitioners for quantitative call center management. 

In this paper we model an inbound telephone call center with balking and abandonment, i.e. the customer may not stay in the queue once realizing he is put on hold, or abandon the queue if the waiting time is too long, extending the nonstationary M/M/n queuing model analyzed by the author in \cite{burak2014multi}.

The paper is structured as follows. In the next section the model and the basic notation are introduced. Section 3 reviews the proposed multi-step uniformization algorithm with steady-state detection and section 4 presents the results of numerical experiments. The paper ends with a summary of results,  conclusions and proposals for future research.

\section{Model}

We propose a following model of a Call Centre: the analyzed period is finite (e.g. one working day) with the system starting empty. The state variable $X(t)$ represents total number of service requests  (served/waiting calls) in the system at time $t$. The size $n(t)$ of the system, which represents the number of non empty possible states, is finite, equal to $s(t)$ = number of identical servers (agents) plus $q(t)$ = capacity of the queue, with corresponding discrete state space $\varphi(t) = \{0,..,n(t)\}$,$|\varphi(t)|=1+s(t)+q(t)$. 
Customers arrive according to an inhomogenous Poisson process with rate $\lambda(t)$, the service time is i.i.d. exponentially distributed with rate $\mu(t)$. The load $\rho(t) = \lambda(t)/s(t) \mu(t)$ can be bigger than 1. 

Service requests that are not served immediately can leave the system (hang up or balk) with probability 1-$\gamma$, otherwise, after joining the queue, they abandon after reaching their \emph{patience time}. The patience times are independent and identically exponentially distributed with mean $1/\eta$. Queued requests are FCFS served. All of this is modeled via the state transition rates of a CTMC which is described by \emph{infinitesimal generator matrix} $Q(t):n(t)+1 \times n(t)+1, Q(t)=(q_{i,j}(t))$
and the \emph{initial state probability vector} $p(0)$,
where the time dependent value $q_{i,j}(i \neq j)$ is the rate at which the state $i$
changes to the state $j$ and
$q_{i,i}=-\sum_{j \neq i} q_{i,j}$
represents the  rate for the event of staying in the same state.

Because $X(t)=k$ is a birth-and-death process, it can be described by following state dependent birth $q_{k,k+1}(t) = \lambda_k(t)$ and death $q_{k,k-1}(t) = \mu_k(t)$ rates:

\begin{eqnarray}
\label{burak_lambda}
\lambda_k(t) & = &
\begin{cases}
\lambda(t),  & \text{if } 0 \leq k \leq s(t)-1\\
\gamma\lambda(t),  & \text{if } s(t) \leq k\leq n-1\\
\end{cases}
\\
\label{burak_mju}
\mu_k(t) & = &
\begin{cases}
k\mu(t), & \text{if } 1 \leq k \leq s(t)-1\\
s(t)\mu(t) + (k-s(t))\eta, & \text{if } s(t) \leq k\leq n\\
\end{cases}
\end{eqnarray}

Similar to the model $M_1$ in \cite{Deslauriers_2007}. The transient distribution at time t $p(t)$ for a given time dependent generator matrix $Q(t)$ can be calculated using Kolmogorov`s forward equations:
\begin{equation}\label{qih} 
p`(t)=p(t)Q(t)
\end{equation}
where the vector $p(t)=[p_{0}(t),...,p_{n}(t)]$ gives probabilities of the system being in any of the states at time $t$.
 
As we do not allow blocking or abandonment due to the overflow of the system, the capacity of the queue has to be big enough to be considered practically infinite, which is insofar realistic, as the cost of setting practically unlimited queue space in the telecommunications equipment is negligible nowadays.
The system size must, in consequence, ensure that the probability of being in the state $n$ (blocking or abandoning service requests) is insignificant compared to the required computational precision of the whole model.

\section{Multi-Step Uniformization with Steady-State Detection}
The infinitesimal generator matrix $Q(t)$ of an inhomogenous continuous-time Markov chain (ICTMC) is time dependent and the process is described by modified Kolmogorov`s forward equations \eqref{qih}.

When the changes in generator matrix Q occur in a discrete way at finite points of time and all rates are constant during the intervals between them, we could also replace the analyzed ICTMC with a sequence of homogeneous systems computing the state probability vectors for consecutive time periods recursively using uniformization as proposed e.g. in \cite{Rindos_1995} or in \cite{Gross_1984}.

In case of a call center, time dependent changes in $Q$ can occur either discretely due to the changing number of servers or due to changes in the arrival rate.
Since the forecast and current traffic data in call center Management applications are already aggregated with their average values by an arbitrary period (e.g. 5, 15 or 30min), we will further assume, similarly to \cite{Ingolfsson_2010}, Q(t) being accordingly piecewise constant and refer to such consecutive time periods of length $\Delta$ with the coresponding homogenous continuous-time Markov chains (HCTMCs) as steps.

Another approach adopting uniformization for time-inhomogenous CTMCs introduced by \cite{van_Dijk_1992} with subsequent improvements by \cite{van1998numerical}, \cite{Arns_2010} and \cite{andreychenko2010fly} could be used if continuous arrival rates were available, reducing the error of the approximation with the average rates.

Uniformization or Randomization, known since the publication of Jensen in 1953 and, therefore, often referenced to as Jensen method, is the method of choice for computing transient behavior of CTMCs. Many authors compared its performance in different applications with the conclusion that it usually outperforms known differential equation solvers (e.g. \cite{Grassmann_1978}, \cite{Reibman_1988}, \cite{Arns_2010}). 
To use uniformization we first define the matrix
\begin{equation}
\label{pmx}
P=I+\frac{Q}{\alpha}
\end{equation} which for $\alpha\geq max_{i}(|q_{i,i}|)$ is a stochastic matrix. The value of $\alpha$ is called uniformization rate. Further, let
\begin{equation}
\beta(\alpha{}t,k)=e^{-\alpha{} t} \frac{(\alpha{}t)^k}{k!}
\end{equation}
be the probability of a Poisson process with rate $\alpha$ to generate $k$ events in the interval $[0,t)$. One now finds for $p(t)$
\begin{equation}
\label{jensen}
p(t)=p(0)\sum_{k=0}^{\infty}\beta(\alpha{}t,k)(P)^k
\end{equation}
The formula (\ref{jensen}) can be interpreted as a discrete time Markov process (DTMC) embedded in a Poisson process generating events at rate $\alpha$.

The implemented uniformization algorithm is based on \cite{Reibman_1988}
and computes transient state probabilities for a CTMC  with the following modification of (\ref{jensen}) :
\begin{equation}
\label{reibmann88}
p(t)=\sum_{i=0}^{\infty}\Pi(i)e^{-\alpha t} \frac{(\alpha t)^{i}}{i!}
\end{equation}
where  $\alpha$ is uniformization rate, as described in (\ref{pmx}), and $\Pi(i)$ is the state probability vector of the underlying DTMC after each step $i$ computed iteratively by:
\begin{equation}
\label{reibmann88pi}
\Pi(0)=p(0),\  \Pi(i)=\Pi(i-1)P
\end{equation}
To compute $p(i)$, within prespecified error tolerance, in finite time, the computation stops when the remaining value of cdf of Poisson distribution is less than the error bound $\epsilon$:
\begin{equation}
\label{reibmann88k}
1-\sum_{i=0}^{k}e^{-\alpha t} \frac{(\alpha t)^{i}}{i!} \leq \epsilon
\end{equation}
with $k$ being the \emph{right truncation point}.
As $\alpha t$ increases, the corresponding probabilities of small number of $i$ Poisson events occurring become less significant. This allows us to start the summation from the $l$`th iteration called \emph{left truncation point} with the equation \ref{reibmann88} reduced to:
\begin{equation}
\label{reibmann88truncated}
p(t)=\sum_{i=l}^{k}\Pi(i)e^{-\alpha t} \frac{(\alpha t)^{i}}{i!}
\end{equation}
 \cite{Reibman_1988} suggests that the values of $l$ and $k$ be derived by:
\begin{equation}
\label{reibmann88lk}
\sum_{i=0}^{l-1}e^{-\alpha t} \frac{(\alpha t)^{i}}{i!} \leq \frac{\epsilon}{2},\ 1-\sum_{i=0}^{k}e^{-\alpha t} \frac{(\alpha t)^{i}}{i!} \leq \frac{\epsilon}{2}
\end{equation}

The main computational effort of the algorithm lies in consecutive $k$  matrix vector multiplications (MVM), necessary for calculation of epochs of DTMC in (\ref{reibmann88pi}), and is of $O(\eta k)$ where $\eta$ is the number of nonzero elements of (sparse) $P$. 
For large $\alpha t$, as the distribution converges to normal, both left and right truncation points $l$ and $k$ in (\ref{reibmann88lk}) will tend to be symmetric to the mean. The number $\frac{l+k}{2}$ is consequently of $O(\alpha t)$ and the number of additional $\frac{k-l}{2}$ MVMs for the given error tolerance  of $O\sqrt{\alpha t}$ and proportional to inverse cdf for that given $\epsilon$.
Therefore, although we could solve the $p(t)$ with any accuracy $\epsilon > 0$, choosing a higher, acceptable for a respective practical application, value would bring some computational advantage.

The savings due to (tighter) left truncation are, however, rather insignificant, unless the computation of the first significant DTMC is performed in a more efficient way. 

An example of this, presented first in \cite{muppala1992numerical}, is based on recognizing the steady-state of the underlying DTMC. If convergence of the probability vector in (\ref{reibmann88pi}) is guaranteed then we can stop the MVM after arriving at the steady-state, i.e. let us assume that DTMC has the steady state solution $ \Pi (\infty )$
and that after the $S$ iteration of (\ref{reibmann88pi}) $\Vert\Pi(S) - \Pi(\infty)\Vert_v < \delta(S)$, where $\Vert.\Vert_v$ is an arbitrary vector norm. Then (\ref{reibmann88truncated}) changes to:
\begin{equation}
\label{muppala92}
\hat{p}(t)=
\begin{cases}
\Pi(S) & \text{if } S\leq l,\\
\displaystyle{\sum_{i=l}^{S}\Pi(i)e^{-\alpha t} \frac{(\alpha t)^{i}}{i!} + \Pi(S)(1 - \sum_{i=0}^{S}e^{-\alpha t} \frac{(\alpha t)^{i}}{i!})}& \text{if } l<S \leq k,\\
\text{same as }p(t) \text{ in }(\ref{reibmann88truncated}) & \text{if } S>k
\end{cases}
\end{equation}
with $\hat{p}(t)$ used instead $p(t)$ denoting transient state probability vector computed using approximate steady state DTMC vector $\Pi(S)$.
According to \cite{Malhotra_1994} for a predefined error bound $\epsilon$ (as in \eqref{reibmann88k},\eqref{reibmann88lk}) the following inequality holds:

\begin{equation}
\label{malhotra94error}
\ \Vert p(t) - \hat{p}(t) \Vert < \frac{\epsilon}{2} + 2\delta (S)
\end{equation}

The computing of consecutive epochs of the DTMC is equivalent to the power method of finding stationary probability vector of a finite Markov chain. According to \cite{stewart2009probability} if the stochastic matrix $P$ is aperiodic convergence of the power method is guaranteed and the number of iterations $k$ needed to satisfy a tolerance criterion $\xi$ may be obtained approximately from the relationship 

\begin{equation}
\label{stewart2009}
\ \rho ^k = \xi  \text{, i.e.,  }  k=\frac{log \xi}{log \rho}
\end{equation}

where $\rho$ is the magnitude of subdominant eigenvalue $\lambda _2$ of matrix $P$ 
\begin{equation}
\label{eigenvals}
1=\Vert \lambda _1 \Vert > \Vert \lambda _2 \Vert \geq \Vert \lambda _3 \Vert ... \geq \Vert \lambda _N \Vert 
\end{equation}
reducing, consequently, the computational complexity to $O(\eta \  log \xi / log \vert \lambda_2 \vert )$. 

Since in most cases the size of the subdominant eigenvalue is not known in advance, the usual method of testing for convergence is to examine some norm of the difference of successive iterates:
\begin{equation}
\label{stewart2009iterates}
\Vert \Pi_i(k) - \Pi_i(k-m) \Vert < \xi
\end{equation}
 \cite{stewart2009probability} recommends using the relative convergence test of iterates spaced apart by $m$ being function of the rate of convergence:

\begin{equation}
\label{stewart2009test}
max_i\left( \frac{\vert \Pi_i(k) - \Pi_i(k-m) \vert}{\vert \Pi_i(k) \vert} \right) < \xi  \end{equation}

and suggests envisaging a "battery" of different convergence tests in order to accept the approximation $\Pi(S)$ as being sufficiently accurate.
The main risk in this approach is that in order to ensure, with the above proposed methods, that the $\Pi(S)$ is steady, an additional computational effort  for both the convergence tests and the required additional number of iterations can easily obliterate the potential savings.

However, in case of our model, we can easily calculate precise stationary distribution $\Pi(\infty)$  in advance, using global balance equations (e.g. as in \cite{stewart2009probability}) with birth and death rates as in \eqref{burak_lambda} and \eqref{burak_mju}.
Therefore, we can consequently, as proposed in \cite{burak2014multi} instead of iterating the DTMC vector in \eqref{reibmann88pi} up to a point $S$ where it would probably satisfy required convergence tests, simply use the  $\Pi(\infty)$ (instead of $ \Pi(S)$, as proposed in the original algorithm by \cite{muppala1992numerical}) as the $\hat{p}(t+\Delta)$ approximation of $p(t+\Delta)$.

This can be decided after relatively few $i$ iterations due to convergence properties of the power method as described  e.g. in \cite{O_Leary_1979} or in standard books on numerical analysis, using numerically estimated convergence function of $\Pi(i)$ (as proposed in \cite{burak2014multi}), as it allows for precise calculation of the error of such a solution:
 \begin{equation}
\label{burak13sserr_eps_taylor}
\varepsilon_{t+\Delta} = \frac{\Vert \Pi(l) - \Pi(\infty) \Vert_\infty}{\Vert \Pi(\infty) \Vert_\infty}
\end{equation}
in order to decide if it is acceptable (smaller than a predefined steady-detection threshold $\delta_t$).

One of the biggest advantages of the uniformization is its strict error bounding for one step independently of its length. It is not difficult to show (e.g. \cite{van_Moorsel_1997}) that the total error for a number of uniformization steps is the sum of truncation errors (error bounds) for each step.

Assume for a time period $T$ with a known initial distribution $p(0)$ that for any $p(\tau)$, $\tau=(0,T]$ the value of each its state has to be computed with an error less than $\varepsilon_T$. Let us further assume $\varepsilon_t < \varepsilon_T$ being the error after computing some $p(t), t<T$. Then:
\begin{equation}
\label{silvaepsilon}
 \varepsilon_t + \sum_i {\epsilon_{\Delta_i}}  \leq \varepsilon_T ,\  \sum_i {\Delta_i} = T-t
\end{equation}

As the error bound of steady state approximation is, in case the steady state is reached, absolute and independent of the error of the previous steps, we can set the convergence threshold dependent rather on the  actual total error bound than the error for the single step (as proposed e.g. by \cite{Malhotra_1994}). It allows, consequently, to trade the error bounds of steps for higher convergence thresholds while still within the global error bound for the whole solution. Then, assuming the system at time $m,0 \leq m <T$ -- to satisfy $\varepsilon_t < \varepsilon_T$ for each $p(t)$, $t=(m,T]$ we have to:

\begin{equation}
\begin{split}
\label{burak13delta}
\delta_m \leq \varepsilon_T - \varepsilon_m - \sum_m^{T}\epsilon_\Delta
\end{split}
\end{equation}

\section{Computational Examples}
\begin{figure}[h!t!b]
\caption{Computational example - System load}
\label{load}
\centering
\begin{minipage}{0.86\textwidth} 
\includegraphics[width=\linewidth]{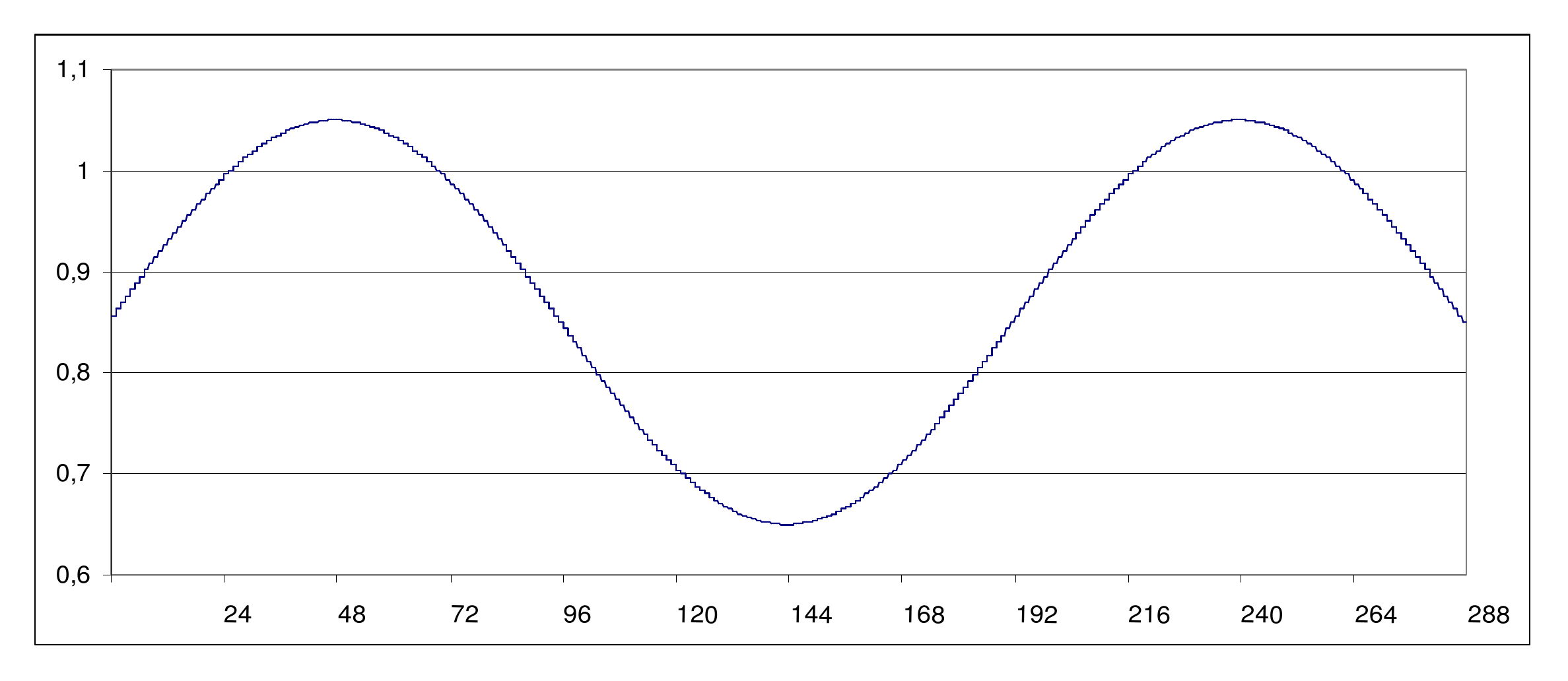}
\end{minipage}
\end{figure}

To test the implementation the following model has been used: a service system (call center) working for time $T$ = 24h and starting empty.  The arrival rate changes sinusoidal with two peaks and is divided into 288 (5min) periods with constant averaged rates, same as the first example in \cite{burak2014multi}. The service rate  and number of servers  are constant ($\mu(t)=\mu$, $s(t)=s$), the arrival rate varies in time - $\lambda (t) = s \mu (0.85 + 0.2sin(3\pi t / T), 0 \leq t < T$ (the load varying between 0.65 and 1.05 as shown in Figure \ref{load}). The probability $1-\gamma$ of a customer immediately leaving when not served immediately is 0.03. The mean value of patience time $1/\eta$ is equal to 4min.

The capacity of the queue is constant and chosen so that for all times the probability $p_n(t)$ of the system being in the state $n$ is less than $1 \times 10^{-5}$ for all tested system sizes.
To evaluate the impact of the proposed steady-state detection algorithm, models of 5 different sizes have been at first calculated  using unmodified uniformization algorithm with an error step $\epsilon=1.5\times 10^{-5}$ corresponding to the total error bound $\varepsilon_T=2.88 \times 10^{-3}$.

\begin{table}[h]
\caption{Computation times, steady-state detection (avx).}
\resizebox{0.8\textwidth}{!}{\begin{minipage}{\textwidth}
{\begin{tabular}{l||r|r||r|r||r|r||r|r||r|r}
$\epsilon_\Delta$=1e-7 & \multicolumn{2}{c||}{$\delta=$0($\epsilon_\Delta$=1e-5)} & \multicolumn{2}{c||}{$\varepsilon_T = $ 5e-03} & \multicolumn{2}{c||}{$\varepsilon_T = $ 1.5e-02} &  \multicolumn{2}{c||}{$\varepsilon_T = $ 3e-02} & \multicolumn{2}{c}{$\varepsilon_T = $ 5e-02} \\
\hline
System size & time & $t/n^2$ & time & $t/n^2$ & time & $t/n^2$ & time & $t/n^2$ & time & $t/n^2$ \\
\hline
54.....(30+24)&		4.25&	1.46&	4.27&	1.46&	3.25&	1.11&	2.13&	0.731&	2.48&	0.850\\
150...(100+50)&		15.8&	0.70&	15.3&	0.68&	11.5&	0.51&	4.46&	0.198&	3.73&	0.166\\
390...(300+90)&		90.1&	0.59&	86.2&	0.57&	62.0&	0.41&	43.6&	0.286&	15.3&	0.101\\
1200(1000+200)&		709&	0.49&	715&	0.50&	534&	0.37&	468&	0.325&	378&	0.262\\
3300(3000+300)&		5996&	0.55&	5547&	0.51&	4350&	0.40&	4053&	0.372&	3915&	0.359\\
\end{tabular}}
{\scriptsize load $0.65 \leq \rho \leq 1.05$\par}
\end{minipage}}
\label{table_ssd_taylor}
\end{table}

The detailed results of computation times are in Table \ref{table_ssd_taylor}.

All experiments were performed on a 1.7GHz PC under 64bit Linux OS with a processor supporting vector operations in both: avx with 256bit vectors (4 double or 8 float operations simultaneously) and the older sse instruction set with 128bit vector operations (an Intel i5-3317U with cpu throttling disabled via kernel scaling governor), compiled with GNU GCC compiler.

All measurements use standard Unix \emph{time.h/clock()} function - returning CPU time. All times are in milliseconds.
\begin{figure}[h!t!b]
\caption{Number of iterations (mvm) per step, system size 1200.}
\label{ssd_abdmnt_65_105_taylor_1000_200}
\centering
\begin{minipage}{0.86\textwidth} 
\includegraphics[width=\linewidth]{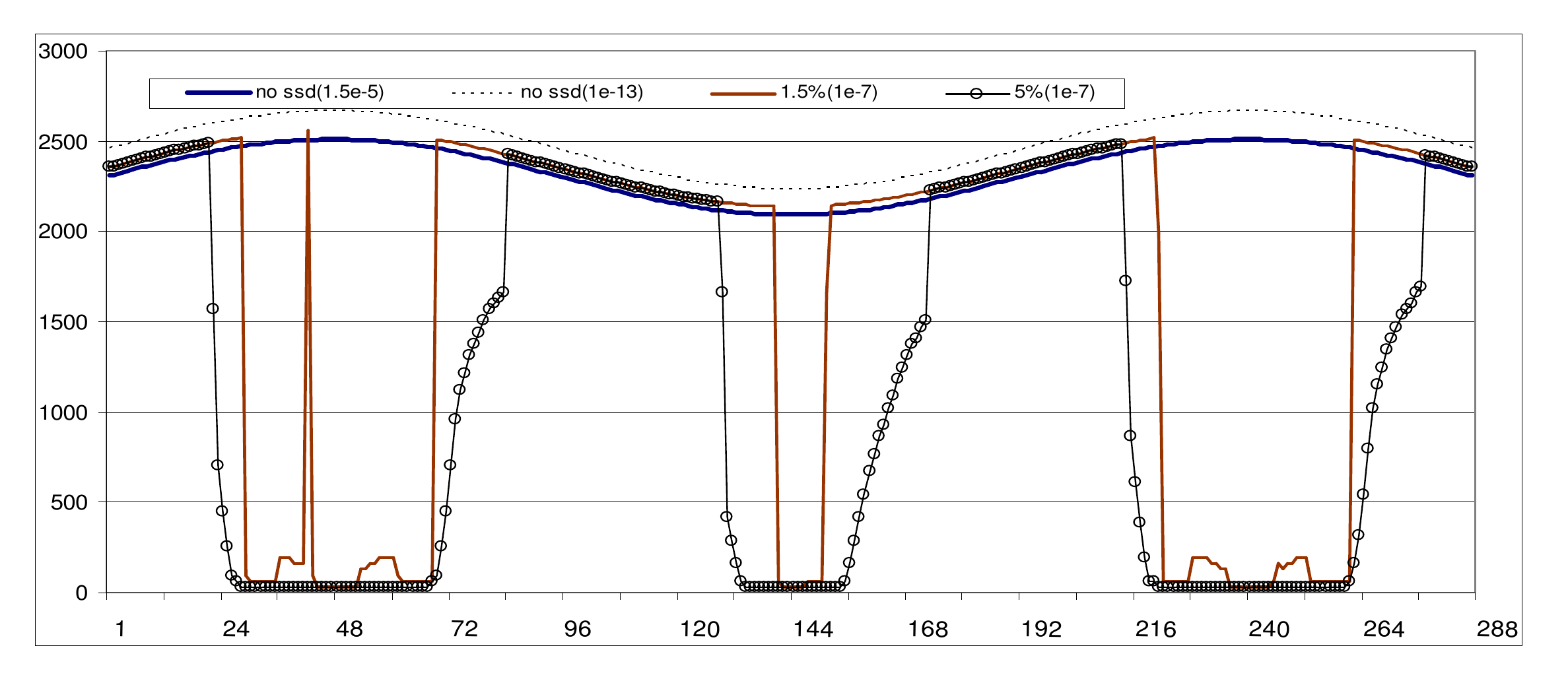}
{\scriptsize load $0.65 \leq \rho \leq 1.05$, s=1000 q=200\par}
\end{minipage}
\end{figure}
The impact of reduced computational effort due to steady-state detection for some chosen total error bounds (between $0$ and $5\times10^{-2}$), with corresponding steady-state detection thresholds, is illustrated for the system of size 1200 in Figure \ref{ssd_abdmnt_65_105_taylor_1000_200}. 

\begin{figure}[h!t!b]
\caption{Expected system state, system size 1200.}
\label{ssd_abdmnt_65_105_ESS_taylor_1000_200}
\centering
\begin{minipage}{0.86\textwidth} 
\includegraphics[width=\linewidth]{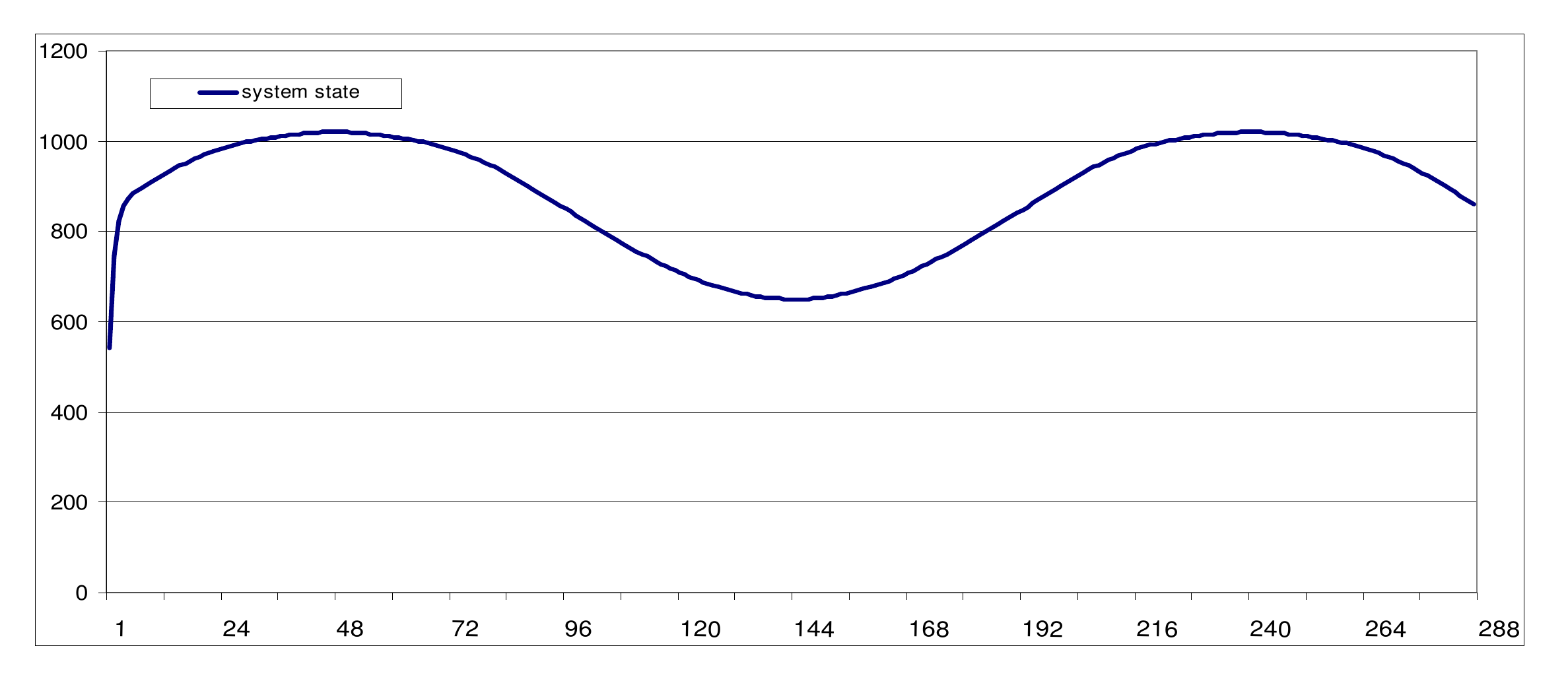}
{\scriptsize load $0.65 \leq \rho \leq 1.05$, s=1000 q=200\par}
\end{minipage}
\end{figure}

\begin{figure}[h!t!b]
\caption{Error of the expected system state, system size 1200.}
\label{ssd_abdmnt_65_105err_taylor_1000_200}
\centering
\begin{minipage}{0.86\textwidth} 
\includegraphics[width=\linewidth]{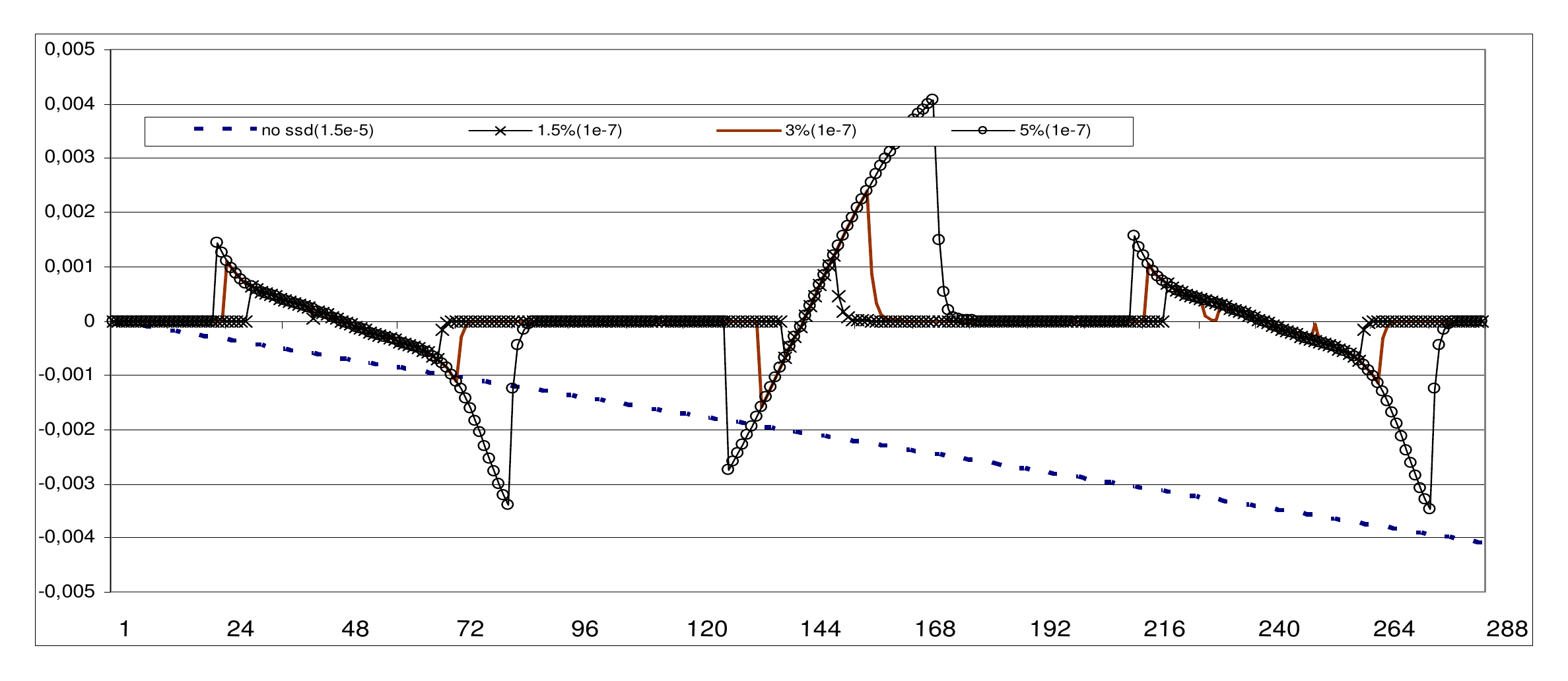}
{\scriptsize load $0.65 \leq \rho \leq 1.05$, s=1000 q=200\par}
\end{minipage}
\end{figure}
Figure \ref{ssd_abdmnt_65_105_ESS_taylor_1000_200} shows the expected state of the system, derived from the calculated probability vector as:

\begin{equation}
\label{burak13es}
ES(t)=\sum_i{i p_i(t)},\  p(t)=[p_0..p_n]
\end{equation}

Figure \ref{ssd_abdmnt_65_105err_taylor_1000_200} shows its relative error for different steady-state detection thresholds. The reference for the error estimate has been calculated with $\epsilon_\Delta=1\times10^{-13}$.

\begin{figure}[h!t!b]
\caption{Probability of a request being served immediately, system size 1200.}
\label{ssd_abdmnt_65_105_SL0_taylor_1000_200}
\centering
\begin{minipage}{0.86\textwidth} 
\includegraphics[width=\linewidth]{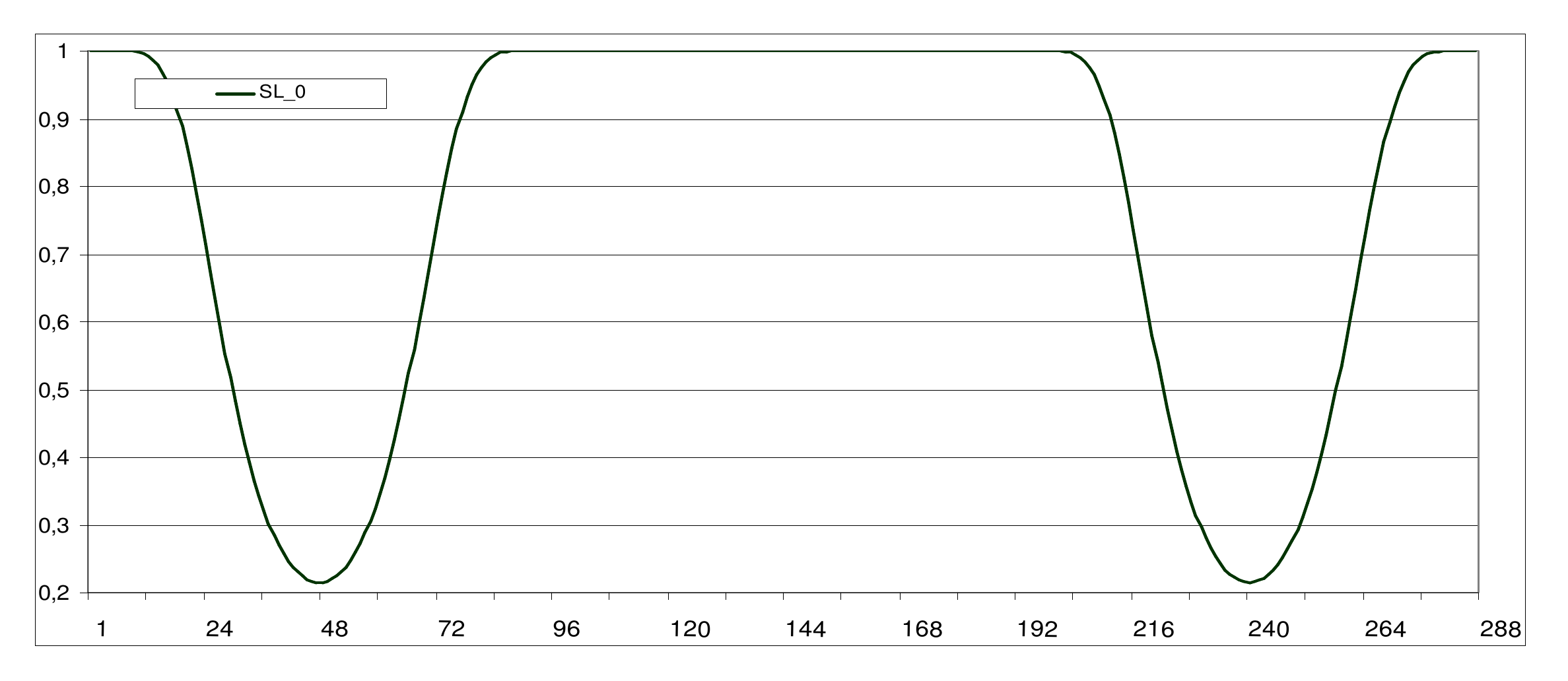}
{\scriptsize load $0.65 \leq \rho \leq 1.05$, s=1000 q=200\par}
\end{minipage}
\end{figure}

Figure \ref{ssd_abdmnt_65_105_SL0_taylor_1000_200} shows the probability for an incoming service request to be served immediately (with no waiting time).

\begin{table}[h]
\caption{Computation time and p(n) depending on queue length, $\gamma$ and $1/\eta$ .}
\resizebox{0.82\textwidth}{!}{\begin{minipage}{\textwidth}
{\begin{tabular}{l||c|l||c|l||c|l}
$\epsilon_\Delta$ = 1e-7, $\varepsilon_T=$ 3e-02& \multicolumn{2}{c||}{$\gamma$=0.97, $1/\eta$= 8min} & \multicolumn{2}{c||}{$\gamma$=0.99, $1/\eta$= 8min} & \multicolumn{2}{c}{$\gamma$=0.997, $1/\eta$=12min} \\
\hline
System size & time & max $ p_n(t)$ & time & max $ p_n(t)$ & time & max $ p_n(t)$ \\
\hline
 & & & & &\\[-4pt]
1250..(1000+250)& 479&	$6.6\times10^{-9}$ &	661& $3.8\times10^{-7}$ &	686& $1.8\times10^{-4}$\\[2pt]
1300..(1000+300)& 530&	$5.8\times10^{-12}$&	715& $9.3\times10^{-10}$&	760& $7.7\times10^{-6}$\\[2pt]
1400..(1000+400)& 607&	$9.2\times10^{-20}$&	827& $1.1\times10^{-16}$&	813& $8.7\times10^{-10}$\\[2pt]
\end{tabular}}
{\scriptsize load $0.65 \leq \rho \leq 1.05$\par}
\end{minipage}}
\label{table_gamma_eta_65_105}
\end{table}

Table \ref{table_gamma_eta_65_105} shows computation times and maximal values of  $p_n(t), 0<t<T$ -- probability of the system being in the state $n$,  for different values of: queue length, probability of the customer entering the queue despite not being served immediately $\gamma$ and mean value of patience time $1/\eta$. The difference compared to the corresponding value for $\gamma=0.97$ and $1/\eta$= 4min from the Table \ref{table_ssd_taylor} (equal to 468ms) is not only both due to the bigger size of the system and higher uniformization rate $\alpha$ resulting from higher queue length, but also to some extent  due to the higher variability of the system state, resulting in fewer steps where the steady-state could be detected within the respective threshold. To illustrate this effect, we repeated the experiment with the load variability reduced to only $0.95<\rho<1.05$ i.e. with the arrival rate  $\lambda (t) = s \mu (1.0 + 0.05sin(3\pi t / T), 0 \leq t < T$. The results corresponding to the cases from the Table \ref{table_gamma_eta_65_105} are shown in the Table \ref{table_gamma_eta_95_105}.

\begin{table}[h]
\caption{Computation time and p(n) depending on queue length, $\gamma$ and $1/\eta$ .}
\resizebox{0.82\textwidth}{!}{\begin{minipage}{\textwidth}
{\begin{tabular}{l||c|l||c|l||c|l}
$\epsilon_\Delta$ = 1e-7, $\varepsilon_T=$ 3e-02& \multicolumn{2}{c||}{$\gamma$=0.97, $1/\eta$= 8min} & \multicolumn{2}{c||}{$\gamma$=0.99, $1/\eta$= 8min} & \multicolumn{2}{c}{$\gamma$=0.997, $1/\eta$=12min} \\
\hline
System size & time & max $ p_n(t)$ & time & max $ p_n(t)$ & time & max $ p_n(t)$ \\
\hline
 & & & & &\\[-4pt]
1250..(1000+250)& 34.6&	$6.6\times10^{-9}$ &	93.6& $3.8\times10^{-7}$ &	320& $1.8\times10^{-4}$\\[2pt]
1300..(1000+300)& 34.4&	$5.8\times10^{-12}$&	98.8& $9.3\times10^{-10}$&	349& $7.7\times10^{-6}$\\[2pt]
1400..(1000+400)& 38.0&	$9.2\times10^{-20}$&	110& $1.1\times10^{-16}$&	380& $8.7\times10^{-10}$\\[2pt]
\end{tabular}}
{\scriptsize load $0.95 \leq \rho \leq 1.05$\par}
\end{minipage}}
\label{table_gamma_eta_95_105}
\end{table}


\section{Conclusion}
In this paper we showed that the uniformization with steady-state detection can be used in a very effective way to evaluate transient behavior of multiserver queues. Applied to the modeling of the call center schedules, it allows calculation of transient system states for systems of any, possible in practical applications, size in a very short time, in a numerically stable way, with very high precision, using relatively common and inexpensive CPU. It can, therefore, be used for schedule planning based on available forecasts, as described in \cite{Ingolfsson_2010}. 

The presented method can be extended in several directions. One could be, in regard to call center modeling, to automatically optimize the model size (queue length) with significant impact on the computational efficiency. Another could be to use known periodicity of traffic forecasts to divide total error bound in between known times of the day, bounded by the points of time when the system will reach a steady state, than for the whole modeled period.

\newpage

\end{document}